# Markovian inhomogeneous closures for Rossby waves and turbulence over topography

**Jorgen S. Frederiksen and Terence J. O'Kane**

CSIRO Oceans and Atmosphere, Aspendale, Victoria and Hobart Tasmania, Australia;
E-Mail: Jorgen.Frederiksen@csiro.au

**Abstract:** Manifestly Markovian closures for the interaction of two-dimensional inhomogeneous turbulent flows with Rossby waves and topography are formulated and compared with large ensembles of direct numerical simulations (DNS) on a generalized $\beta$– plane. Three versions of the Markovian inhomogeneous closure (MIC) are established from the quasi-diagonal direct interaction approximation (QDIA) theory by modifying the response function to a Markovian form and employing respectively the current-time (quasi-stationary) fluctuation dissipation theorem (FDT), the prior-time (non-stationary) FDT and the correlation FDT. Markov equations for the triad relaxation functions are derived that carry similar information to the time-history integrals of the non-Markovian QDIA closure but become relatively more efficient for long integrations. Far from equilibrium processes are studied, where the impact of a westerly mean flow on a conical mountain generates large amplitude Rossby waves in a turbulent environment, over a period of 10 days. Excellent agreement between the evolved mean streamfunction and mean and transient kinetic energy spectra are found for the three versions of the MIC and two variants of the non-Markovian QDIA compared with an ensemble of 1800 DNS. In all cases mean Rossby wavetrain pattern correlations between the closures and the DNS ensemble are greater than 0.9998.

## 1. Introduction

Modern statistical dynamical closure theory, initially applied to the iconic problem of homogeneous isotropic turbulence (McComb 2014), has its origin in the pioneering works of Kraichnan (1959a) who derived the equations for his Eulerian direct interaction approximation (DIA) closure on the basis of formal renormalized perturbation theory. His approach had some elements in common with the renormalized perturbation theory and functional approaches to quantum electrodynamics (QED) developed in the mid-20[th] century by Tomonaga, Schwinger and Feynman (Frederiksen 2017 reviews the literature). Unlike QED, where the fine structure constant, measuring interaction strength, is only $\sim 1/137$, turbulence at high Reynolds number is a problem of strong interaction. The DIA is a two-point non-Markovian closure for the renormalized two-time covariances or cumulants and response functions but the interaction coefficients – vertices –are unrenornalized or bare. Herring (1965, 1966) subsequently developed his self-consistent field theory (SCFT) closure and McComb (1974, 1990, 2014) developed his local energy transfer (LET) closure through independent approaches. These two-point non-Markovian closures were later shown to differ from the DIA only in how a fluctuation-dissipation theorem (FDT) (Kraichnan 1959b; Deker & Haake 1975) is invoked (Frederiksen et al., 1994; Kiyani & McComb 2004). The *prior-time FDT* (Carnevale & Frederiksen 1983a, equation (3.5)) relates the two-time spectral covariance $C_\mathbf{k}(t,t')$ at wavenumber **k** to the response function $R_\mathbf{k}(t,t')$ and the prior single-time covariance $C_\mathbf{k}(t',t')$ through

$$C_\mathbf{k}(t,t') \equiv R_\mathbf{k}(t,t') C_\mathbf{k}(t',t') \tag{1.1}$$

for $t \geq t'$ as discussed in more detail in Section 5. The SCFT has the same statistical dynamical equations for the single-time covariance and response function as the DIA but replaces the two-time covariance with the approximate expression in the FDT. The LET on the other hand has the same equations for the single-time and two-time covariances as the DIA but uses equation (1.1) to determine the response function.

The Eulerian DIA and the SCFT and LET closures result in energy spectra that generally agree well with the statistics of DNS in the energy containing range of the larger scales. However, at high Reynolds numbers, the DIA power laws differ slightly from the classical $k^{-5/3}$ energy and $k^{-3}$ enstrophy cascading inertial ranges. These discrepancies have been attributed to the fact that the DIA does not distinguish adequately between sweeping effects and intrinsic distortion effects resulting in spurious non-local interactions between the large and small scale eddies (Kraichnan 1964a; Herring et al. 1974; Frederiksen & Davies 2000 review the literature). In order to overcome these power law deficiencies, quasi-Lagrangian versions of the direct interaction theories were developed by Kraichnan (Kraichnan 1965, 1977; Kraichnan & Herring 1978) and Kaneda (Kaneda 1981; Gotoh et al. 1988). The non-Markovian quasi-Lagrangian theories, in common with the Eulerian DIA, contain no ad hoc parameters but the results depend on whether the closure is formulated in terms of labeling time derivatives, like Kraichnan's Lagrangian-history direct interaction (LHDI) or in terms of measuring time derivatives, like Kaneda's Lagrangian renormalized approximation (LRA) and also on an ad hoc choice of the basic variables used (Kraichnan 1977; Kaneda 1981; Herring & Kraichnan 1979; Frederiksen & Davies 2004 review the literature). Interestingly, the Eulerian LET closure of McComb (1974, 1990), is also consistent with the Kolmogorov (1941) classical $k^{-5/3}$ inertial range for high Reynolds number three-dimensional turbulence. As well, in contrast to the quasi-Lagrangian closures, the LET, like the Eulerian DIA, is independent of the choice of basic variables such as velocity, vorticity and strain. At finite resolution and moderate Reynolds numbers the performance of the Eulerian DIA, SCFT and LET closures are quite similar and in two dimensions all tend to underestimate the amplitudes of the small scale energy spectra (Frederiksen & Davies 2000; 2004).

The Eulerian and quasi-Lagrangian closures describe the evolution of the renormalized 'propagators', the response functions and two-point cumulants, but have deficiencies that arise from not accounting systematically for "vertex" renormalization, the modification of the strength and form of the interaction coefficients. As noted by Martin et al. (1973) "the whole problem of strong turbulence is contained in a proper treatment of vertex renormalization". This is an unsolved problem for strongly interacting fields in general. However, Kraichnan (1964c) proposed removing some of the convection effects of the large scales on the small scales so that the modified Eulerian DIA becomes consistent with the Kolmogorov inertial range spectra. He zeroed the interaction coefficients – vertex functions – to localize the interactions between triads of wavenumbers dependent on a cut-off ratio $\alpha$ (our Appendix C has further details). Interestingly, we have found that the value of $\alpha$ that gives close agreement at all scales of the energy spectra between ensembles of DNS and these regularized closures is essentially universal (Frederiksen & Davies 2004; O'Kane & Frederiksen 2004, hereafter OF04).

Non-Markovian closures with potentially long time-history integrals present a significant computational challenge, particularly at high resolution, and even more so if the general inhomogeneous problem is attempted. Orszag (1970) employed a heuristic approach to derive a simpler Markovian closure, denoted the eddy damped quasi-normal Markovian (EDQNM) closure, for the evolution of the single-time covariance associated with homogeneous isotropic turbulence. He recognized that in order to reproduce the Kolmogorov inertial range at small scales the molecular viscosity should be augmented by an eddy viscosity that has a functional form consistent with the Kolmogorov power law and a strength determined by an empirical constant. Interestingly, the EDQNM closure for homogeneous isotropic turbulence satisfies an H-theorem that guarantees the monotonic increase of entropy for the inviscid system and approach to a canonical equilibrium solution for two and three-dimensional systems (Carnevale et al. 1981). The EDQNM closure can also be derived by modifying the Eulerian DIA; firstly the response function is replaced by a Markovian form with this eddy viscosity and secondly the two-time covariance is determined from the *current-time FDT*

$$C_{\mathbf{k}}(t,t') \equiv R_{\mathbf{k}}(t,t')C_{\mathbf{k}}(t,t) \qquad (1.2)$$

for $t \geq t'$ (see Section 5). Another one parameter Markovian closure, the test-field model (TFM) was developed by Kraichnan (1971) to yield the Kolmogorov inertial range in three-dimensional turbulence and by Leith & Kraichnan (1972) for two-dimensional turbulence.

Leith (1971) pioneered the numerical implementation of the EDQNM closure for two-dimensional turbulence with application to atmospheric predictability and subgrid modelling. Subsequently the EDQNM has been widely applied to both isotropic and anisotropic problems in two-dimensional and three-dimensional turbulence and the interaction of waves and turbulence (Carnevale & Martin 1982; Carnevale & Frederiksen 1983b; Bowman et al. 1993; Frederiksen & Davies 1997; Cambon et al. 2017 review the literature). Leith & Kraichnan (1972) implemented and employed the TFM closure for studying error growth in two-dimensional and three-dimensional turbulent flows. Herring (1977) and Holloway (1978) applied generalizations of the TFM for studying turbulent homogeneous flows over random topography with zero mean. The TFM has also been applied in many other studies of both isotropic and anisotropic turbulence (Bowman & Krommes 1997 review the literature).

Bowman et al. (1993) showed that for anisotropic turbulence in the presence of linear wave phenomena EDQNM type closures may be potentially nonrealizable. They demonstrated that this was a possibility if the EDQNM was derived as a modification of the DIA closure with Markovian response functions and the two-time covariances determined by the current-time FDT in equation (1.2) or the prior-time FDT in equation (1.1). However, they established a realizable Markovian closure (RMC) by using a response function with positive damping and specifying the two-time covariance through the *correlation FDT*

$$C_{\mathbf{k}}(t,t') \equiv [C_{\mathbf{k}}(t,t)]^{\frac{1}{2}} R_{\mathbf{k}}(t,t') [C_{\mathbf{k}}(t',t')]^{\frac{1}{2}} \quad (1.3)$$

for $t \geq t'$ (see Section 5). That is, equation (1.3) states that the correlation function $C_{\mathbf{k}}(t,t')[C_{\mathbf{k}}(t,t)]^{-\frac{1}{2}}[C_{\mathbf{k}}(t',t')]^{-\frac{1}{2}}$ is equal to the response function $R_{\mathbf{k}}(t,t')$. They determined Markovian equations for the triad relaxation functions rather than specifying empirical analytical forms and also considered multi-field versions. Bowman & Krommes (1997) also developed a realizable test-field theory (RTFM) closure for anisotropic turbulence and applied it and the RMC to study turbulence in the presence of plasma drift waves through the Hasegawa-Mima equation while Hu et al. (1997) performed similar studies for the two-field Hasegawa-Wakatani equations (Krommes 2002 reviews the plasma physics literature).

The aim of this article is to formulate and apply three versions of the Markovian inhomogeneous closure (MIC) to the problem of general two-dimensional inhomogeneous turbulent flows interacting with Rossby waves and topography on a generalized $\beta$– plane. In doing so we extend the EDQNM and RMC closures for homogeneous turbulence to Markovian closures for the general problem of the interaction of inhomogeneous turbulent flows with waves and orography. Three versions of the MIC are obtained from the quasi-diagonal direct interaction approximation (QDIA) theory for inhomogeneous flows by modifying the response function to a Markovian form and replacing the two-time covariance by the three forms of the FDT in equations (1.1) – (1.3) respectively.

The QDIA closure was developed for two-dimensional inhomogeneous turbulent flows over topography on an $f$– plane by Frederiksen (1999, hereafter F99) and generalized to include non-Gaussian and inhomogeneous initial conditions by O'Kane and Frederiksen (OF04) and to Rossby wave turbulence on a $\beta$– plane by Frederiksen & O'Kane (2005, hereafter FO05). It was subsequently formulated for general classical field theories with first order time-derivatives (Frederiksen 2012a, b) and for classical and quantum field theories with first or second order time-derivatives and non-Gaussian noise and non-Gaussian initial conditions (Frederiksen 2017). For inviscid flows the QDIA closure relaxes to a canonical equilibrium solution (Carnevale & Frederiksen 1987) as shown in F99.

The QDIA was numerically implemented for turbulent flows over topography on an $f$– plane (OF04) and $\beta$– plane (FO05). It was shown to be only a few times more computationally demanding than the homogeneous DIA (Frederiksen & Davies 2000, 2004) unlike Kraichnan's (1964b, 1972) inhomogeneous DIA (IDIA) which has not been computed for turbulent fluids. At moderate Reynolds numbers the QDIA compares very favourably with large ensembles of DNS (FO05) while at high Reynolds numbers a regularized version (Appendix C) of the QDIA yields the right small scale power law behaviour (OF04) as for homogeneous turbulence (Frederiksen & Davies 2004). The computational efficiency of the QDIA has also been enhanced through a cumulant update restart procedure (Rose 1985; Frederiksen et al. 1994) that uses non-Gaussian terms in the initial conditions (OF04; FO05). This variant is termed the cumulant update QDIA (CUQDIA). The QDIA closure has been comprehensively tested against large ensembles of DNS for problems in predictability (FO05; O'Kane & Frederiksen 2008a), data assimilation using Kalman and related filters (O'Kane & Frederiksen 2008b, 2010) and subgrid scale parameterizations (O'Kane & Frederiksen 2008c; Frederiksen & O'Kane 2008). The QDIA theory has also been the framework for developing accurate subgrid scale parameterizations for atmospheric and oceanic turbulent flows within quasi-geostrophic and primitive equation models and for three-dimensional boundary layer turbulence in channels as reviewed by Frederiksen et al. (2017).

The paper is structured as follows. In section 2, we summarize the equations for general two-dimensional flows over topography on a generalized $\beta$-plane with doubly periodic boundary conditions. The flows consist of a large-scale component with a zonal velocity that satisfies the form-drag equation and a spectrum of smaller scales that satisfy the barotropic vorticity equation. The corresponding spectral equations are documented in Section 3. The QDIA closure model is presented in Section 4 and the derivation of three versions of the MIC is formulated in Section 5 based on a Markovian form of the response function and using the three FDT relations in equations (1.1) – (1.3) respectively for the two-time covariance. The performance of the three versions of the MIC, compared with the two variants of the non-Markovian QDIA (with and without cumulant update restarts) and an ensemble of 1800 DNS, is described in Section 6. The significance of our findings and conclusions and suggested future studies are presented in Section 7. Appendix A documents the interaction coefficients used in the spectral equations, Appendix B lists relationships between off-diagonal and diagonal elements of the two-point and three-point cumulants and response functions needed for establishing the closures, and Appendix C outlines the steps in developing regularized variants of the MIC.

## 2. Two-dimensional flow over topography on a generalized $\beta$-plane

For this study of the development and performance of three versions of the MIC model we employ the same equations for two-dimensional flow over topography on a generalized $\beta$-plane as introduced by FO05. Again the streamfunction is written in the form $\Psi = \psi - Uy$ where the 'small scales' are determined by $\psi$ and the large-scale westerly flow by $U$.

### A. *Barotropic vorticity equation for the small scales*

The 'small scales' evolve according to the barotropic vorticity equation in the presence of the large-scale flow and topography

$$\frac{\partial \zeta}{\partial t} = -J(\psi - Uy, \zeta + h + \beta y + k_0^2 Uy) + \hat{\nu}\nabla^2 \zeta + f^0. \tag{2.1a}$$

Here, $\zeta$ is the vorticity, $h$ the scaled topography, $\hat{\nu}$ is the viscosity, $f^0$ a forcing function, $\beta$ is the beta effect, and $k_0$ is a wavenumber that, on a sphere, would determine the strength of the vorticity of the solid-body rotation. This wavenumber can be made as small as desirable to recover the standard $\beta$– plane equations. However, the advantage of retaining a small $k_0$ is that it allows us to combine the spectral equations for the small and large scales into a single elegant form. As shown in FO05, there is then a one-to-one correspondence between the spherical geometry and $\beta$ – plane equations that extends to the statistical mechanics equilibrium solutions in the two geometries. The Jacobian is

$$J(\psi,\zeta) = \frac{\partial \psi}{\partial x}\frac{\partial \zeta}{\partial y} - \frac{\partial \psi}{\partial y}\frac{\partial \zeta}{\partial x} \tag{2.1b}$$

and the vorticity is related to the streamfunction through

$$\zeta = \nabla^2 \psi \equiv \left(\frac{\partial^2}{\partial x^2} + \frac{\partial^2}{\partial y^2}\right)\psi. \tag{2.1c}$$

### B. *Large-scale flow equation*

The large-scale flow $U$ evolves according to the form-drag equation

$$\frac{\partial U}{\partial t} = \frac{1}{(2\pi)^2}\int_0^{2\pi} d^2\mathbf{x}\, h(\mathbf{x})\frac{\partial \psi(\mathbf{x})}{\partial x} + \alpha_U(\overline{U} - U). \tag{2.2}$$

The integrations of equations (2.1) and (2.2) are carried out for flows on the doubly periodic plane $0 \le x \le 2\pi, 0 \le y \le 2\pi$. Here $\mathbf{x} = (x,y)$ and the flow $U$ is forced by relaxing it towards $\overline{U}$ with relaxation coefficient $\alpha_U$.

## 3. Spectral equations

Spectral equations corresponding to the system in Section 2 are derived by expanding each of the 'small-scale' functions in a Fourier series. For example,

$$\zeta(\mathbf{x},t) = \sum_{\mathbf{k}\in\mathbf{R}} \zeta_\mathbf{k}(t)\exp(i\mathbf{k}.\mathbf{x}) \tag{3.1a}$$

where

$$\zeta_\mathbf{k}(t) = \frac{1}{(2\pi)^2}\int_o^{2\pi} d^2\mathbf{x}\,\zeta(\mathbf{x},t)\exp(-i\mathbf{k}.\mathbf{x}), \tag{3.1b}$$

$\mathbf{x} = (x,y), \mathbf{k} = (k_x, k_y), k = (k_x^2 + k_y^2)^{1/2}, \zeta_{-\mathbf{k}} = \zeta_\mathbf{k}^*$ and $\mathbf{R}$ is a circular domain in wavenumber space excluding the origin $\mathbf{0}$. As noted in FO05, it is possible to combine the spectral equations for the 'small scales' with the spectral representation of the form drag equation by defining

$$\zeta_{-\mathbf{0}} = ik_0 U, \zeta_\mathbf{0} = \zeta_{-\mathbf{0}}^*, \tag{3.2}$$

as the zero wavenumber spectral component. As well, the interaction coefficients $A(\mathbf{k},\mathbf{p},\mathbf{q})$ and $K(\mathbf{k},\mathbf{p},\mathbf{q})$ need to be generalized as summarized in Appendix A. Thus the spectral form of the vorticity equation may be written as in the same form as for flows on a non-rotating domain (F99; OF04)

$$\left(\frac{\partial}{\partial t}+\nu_0(\mathbf{k})k^2\right)\zeta_\mathbf{k}(t)=\sum_{\mathbf{p}\in\mathbf{T}}\sum_{\mathbf{q}\in\mathbf{T}}\delta(\mathbf{k},\mathbf{p},\mathbf{q})\left[K(\mathbf{k},\mathbf{p},\mathbf{q})\zeta_{-\mathbf{p}}\zeta_{-\mathbf{q}}+A(\mathbf{k},\mathbf{p},\mathbf{q})\zeta_{-\mathbf{p}}h_{-\mathbf{q}}\right]+f_\mathbf{k}^0 \tag{3.3}$$

where $\mathbf{T}=\mathbf{R}\cup\mathbf{0}$. From equation (A.1c), $\delta(\mathbf{k},\mathbf{p},\mathbf{q})=1$ if $\mathbf{k}+\mathbf{p}+\mathbf{q}=0$ and otherwise is 0. Also

$$\nu_0(\mathbf{k})k^2=\hat{\nu}k^2+i\omega_\mathbf{k} \tag{3.4}$$

and the Rossby wave frequency

$$\omega_\mathbf{k}=-\frac{\beta k_x}{k^2}. \tag{3.5}$$

The $\mathbf{k}=\mathbf{0}$ components of $f_\mathbf{k}^0$ and $\nu_0(\mathbf{k})$ are defined by

$$f_\mathbf{0}^0=\alpha_U\bar{\zeta}_\mathbf{0}, \tag{3.6}$$

$$\nu_0(\mathbf{0})k_0^2=\alpha_U. \tag{3.7}$$

## 4. Quasi-diagonal DIA closure equations

We consider next an ensemble of flows satisfying equation (3.3) where the ensemble mean is denoted by $<\zeta_\mathbf{k}>$ and angle brackets denote expectation value. Then we can express the vorticity component for a given realization by

$$\zeta_\mathbf{k}=<\zeta_\mathbf{k}>+\tilde{\zeta}_\mathbf{k} \tag{4.1}$$

where $\tilde{\zeta}_\mathbf{k}$ denotes the deviation from the ensemble mean. The spectral equations (3.3) can then be expressed in terms of $<\zeta_\mathbf{k}>$ and $\tilde{\zeta}_\mathbf{k}$ as follows

$$\left(\frac{\partial}{\partial t}+\nu_0(\mathbf{k})k^2\right)<\zeta_\mathbf{k}>=\sum_\mathbf{p}\sum_\mathbf{q}\delta(\mathbf{k},\mathbf{p},\mathbf{q})[K(\mathbf{k},\mathbf{p},\mathbf{q})\{<\zeta_{-\mathbf{p}}><\zeta_{-\mathbf{q}}>+C_{-\mathbf{p},-\mathbf{q}}(t,t)\}$$
$$+A(\mathbf{k},\mathbf{p},\mathbf{q})<\zeta_{-\mathbf{p}}>h_{-\mathbf{q}}]+\bar{f}_\mathbf{k}^0, \tag{4.2a}$$

and

$$\left(\frac{\partial}{\partial t}+\nu_0(\mathbf{k})k^2\right)\tilde{\zeta}_\mathbf{k}=\sum_\mathbf{p}\sum_\mathbf{q}\delta(\mathbf{k},\mathbf{p},\mathbf{q})[K(\mathbf{k},\mathbf{p},\mathbf{q})\{<\zeta_{-\mathbf{p}}>\tilde{\zeta}_{-\mathbf{q}}+\tilde{\zeta}_{-\mathbf{p}}<\zeta_{-\mathbf{q}}>$$
$$+\tilde{\zeta}_{-\mathbf{p}}\tilde{\zeta}_{-\mathbf{q}}-C_{-\mathbf{p},-\mathbf{q}}(t,t)\}+A(\mathbf{k},\mathbf{p},\mathbf{q})\tilde{\zeta}_{-\mathbf{p}}h_{-\mathbf{q}}]+\tilde{f}_\mathbf{k}^0. \tag{4.2b}$$

Here, and in subsequent equations, the wave vectors lie in the larger $\mathbf{T}$ domain and

$$f_\mathbf{k}^0=\bar{f}_\mathbf{k}^0+\tilde{f}_\mathbf{k}^0, \tag{4.3a}$$

with

$$\bar{f}_\mathbf{k}^0=<f_\mathbf{k}^0> \tag{4.3b}$$

and

$$C_{-\mathbf{p},-\mathbf{q}}(t,s)=<\tilde{\zeta}_{-\mathbf{p}}(t)\tilde{\zeta}_{-\mathbf{q}}(s)>, \tag{4.3c}$$

are two-time covariance (and also two-point cumulant) matrix elements.

We briefly outline the derivation of the QDIA closure equations for barotropic flow over topography by employing the expressions for the off-diagonal elements of the two- and three-point cumulants and response functions, in terms of the diagonal elements for the QDIA closure equations, as described in Appendix B. Firstly, in order to close equation (4.2a) we need the expression for the two-point cumulant $C_{-\mathbf{p},-\mathbf{q}}(t,t)$ in equation (B.1). This results in the expression

$$\sum_{\mathbf{p}}\sum_{\mathbf{q}} \delta(\mathbf{k},\mathbf{p},\mathbf{q}) K(\mathbf{k},\mathbf{p},\mathbf{q}) C_{-\mathbf{p},-\mathbf{q}}(t,t)$$
$$= -\int_{t_o}^{t} ds\, \eta_{\mathbf{k}}(t,s) <\zeta_{\mathbf{k}}(s)> + f_{\mathbf{k}}^{h}(t) \tag{4.4}$$

where
$$\eta_{\mathbf{k}}(t,s) = -4 \sum_{\mathbf{p}}\sum_{\mathbf{q}} \delta(\mathbf{k},\mathbf{p},\mathbf{q}) K(\mathbf{k},\mathbf{p},\mathbf{q}) K(-\mathbf{p},-\mathbf{q},-\mathbf{k}) R_{-\mathbf{p}}(t,s) C_{-\mathbf{q}}(t,s), \tag{4.5a}$$

and
$$f_{\mathbf{k}}^{h}(t) \equiv f_{\mathbf{k}}^{\chi}(t) = h_{\mathbf{k}} \int_{t_o}^{t} ds\, \chi_{\mathbf{k}}(t,s), \tag{4.5b}$$

with
$$\chi_{\mathbf{k}}(t,s) = 2\sum_{\mathbf{p}}\sum_{\mathbf{q}} \delta(\mathbf{k},\mathbf{p},\mathbf{q}) K(\mathbf{k},\mathbf{p},\mathbf{q}) A(-\mathbf{p},-\mathbf{q},-\mathbf{k}) R_{-\mathbf{p}}(t,s) C_{-\mathbf{q}}(t,s). \tag{4.5c}$$

Here, $\eta_{\mathbf{k}}(t,s)$ is the nonlinear damping, $\chi_{\mathbf{k}}(t,s)$ is a measure of the strength of the interaction of the transient eddies with the topography and $f_{\mathbf{k}}^{h}(t)$ is the eddy-topographic force. The response function is defined for $t \geq t'$ through

$$\tilde{R}_{\mathbf{k},\mathbf{l}}(t,t') = \frac{\delta \tilde{\zeta}_{\mathbf{k}}(t)}{\delta \tilde{f}_{\mathbf{l}}^{0}(t')}, \tag{4.6a}$$

$$R_{\mathbf{k},\mathbf{l}}(t,t') = \left\langle \tilde{R}_{\mathbf{k},\mathbf{l}}(t,t') \right\rangle. \tag{4.6b}$$

In equation (4.5) we have also used the shortened notation

$$R_{\mathbf{k}}(t,t') \equiv R_{\mathbf{k},\mathbf{k}}(t,t'), \tag{4.6c}$$

$$C_{\mathbf{k}}(t,t') \equiv C_{\mathbf{k},-\mathbf{k}}(t,t'). \tag{4.6d}$$

Substituting equation (4.4) into equation (4.2a) yields

$$\left(\frac{\partial}{\partial t}+\nu_0(\mathbf{k})k^2\right)<\zeta_\mathbf{k}>=\sum_\mathbf{p}\sum_\mathbf{q}\delta(\mathbf{k},\mathbf{p},\mathbf{q})A(\mathbf{k},\mathbf{p},\mathbf{q})<\zeta_{-\mathbf{p}}(t)>h_{-\mathbf{q}}$$
$$+\sum_\mathbf{p}\sum_\mathbf{q}\delta(\mathbf{k},\mathbf{p},\mathbf{q})K(\mathbf{k},\mathbf{p},\mathbf{q})<\zeta_{-\mathbf{p}}(t)><\zeta_{-\mathbf{q}}(t)>$$
$$-\int_{t_o}^{t}ds\eta_\mathbf{k}(t,s)<\zeta_\mathbf{k}(s)>+f_\mathbf{k}^h(t)+\bar{f}_\mathbf{k}^0(t). \quad (4.7)$$

From equation (4.2b), we can obtain an equation for the diagonal two-time cumulant, needed in equation (4.5), by multiplying by $\tilde{\zeta}_{-\mathbf{k}}(t')$. Then taking the expectation value we have

$$\left(\frac{\partial}{\partial t}+\nu_0(\mathbf{k})k^2\right)C_\mathbf{k}(t,t')$$
$$=\sum_\mathbf{p}\sum_\mathbf{q}\delta(\mathbf{k},\mathbf{p},\mathbf{q})A(\mathbf{k},\mathbf{p},\mathbf{q})C_{-\mathbf{p},-\mathbf{k}}(t,t')h_{-\mathbf{q}}$$
$$+\sum_\mathbf{p}\sum_\mathbf{q}\delta(\mathbf{k},\mathbf{p},\mathbf{q})K(\mathbf{k},\mathbf{p},\mathbf{q})[<\zeta_{-\mathbf{p}}(t)>C_{-\mathbf{q},-\mathbf{k}}(t,t')$$
$$+C_{-\mathbf{p},-\mathbf{k}}(t,t')<\zeta_{-\mathbf{q}}(t)>+<\tilde{\zeta}_{-\mathbf{p}}(t)\tilde{\zeta}_{-\mathbf{q}}(t)\tilde{\zeta}_{-\mathbf{k}}(t')>]+\int_{t_0}^{t'}dsF_\mathbf{k}^0(t,s)R_{-\mathbf{k}}(t',s) \quad (4.8)$$

where $t>t'$ and $C_\mathbf{k}(t,t')=C_{-\mathbf{k}}(t',t)$ for $t'>t$.

Again, equation (B.1) can be used to express the off-diagonal elements of the two-time cumulant in terms of the diagonal elements. In the quasi-diagonal approximation, the treatment of the three-point cumulant in equation (4.8) is the same as in the standard DIA closure for homogeneous turbulence (Kraichnan, 1959a; Frederiksen et al., 1994; Frederiksen, 2003 contains a simple derivation) and is given in equation (B.3). Thus,

$$\left(\frac{\partial}{\partial t}+\nu_0(\mathbf{k})k^2\right)C_\mathbf{k}(t,t')=\int_{t_0}^{t'}ds(S_\mathbf{k}(t,s)+P_\mathbf{k}(t,s)+F_\mathbf{k}^0(t,s))R_{-\mathbf{k}}(t',s)$$
$$-\int_{t_0}^{t}ds(\eta_\mathbf{k}(t,s)+\pi_\mathbf{k}(t,s))C_{-\mathbf{k}}(t',s). \quad (4.9)$$

In equation (4.9),
$$F_\mathbf{k}^0(t,s)=<\tilde{f}_\mathbf{k}^0(t)\tilde{f}_\mathbf{k}^{0*}(s)>, \quad (4.10a)$$
$$S_\mathbf{k}(t,s)=2\sum_\mathbf{p}\sum_\mathbf{q}\delta(\mathbf{k},\mathbf{p},\mathbf{q})K(\mathbf{k},\mathbf{p},\mathbf{q})K(-\mathbf{k},-\mathbf{p},-\mathbf{q})C_{-\mathbf{p}}(t,s)C_{-\mathbf{q}}(t,s), \quad (4.10b)$$
$$P_\mathbf{k}(t,s)=\sum_\mathbf{p}\sum_\mathbf{q}\delta(\mathbf{k},\mathbf{p},\mathbf{q})C_{-\mathbf{p}}(t,s)[2K(\mathbf{k},\mathbf{p},\mathbf{q})<\zeta_{-\mathbf{q}}(t)>+A(\mathbf{k},\mathbf{p},\mathbf{q})h_{-\mathbf{q}}]$$
$$\times[2K(-\mathbf{k},-\mathbf{p},-\mathbf{q})<\zeta_\mathbf{q}(s)>+A(-\mathbf{k},-\mathbf{p},-\mathbf{q})h_\mathbf{q}], \quad (4.10c)$$

$$\pi_{\mathbf{k}}(t,s) = -\sum_{\mathbf{p}}\sum_{\mathbf{q}}\delta(\mathbf{k},\mathbf{p},\mathbf{q})R_{-\mathbf{p}}(t,s)[2K(\mathbf{k},\mathbf{p},\mathbf{q})<\zeta_{-\mathbf{q}}(t)>+A(\mathbf{k},\mathbf{p},\mathbf{q})h_{-\mathbf{q}}]$$
$$\times[2K(-\mathbf{p},-\mathbf{k},-\mathbf{q})<\zeta_{\mathbf{q}}(s)>+A(-\mathbf{p},-\mathbf{k},-\mathbf{q})h_{\mathbf{q}}] \quad (4.10d)$$

and $\eta_{\mathbf{k}}(t,s)$ is given in equation (4.5a). The additional nonlinear damping term $\pi_{\mathbf{k}}(t,s)$ is a measure of the interaction of transient eddies with the mean flow and topography. Both the nonlinear noise terms, $S_{\mathbf{k}}(t,s)$, due to eddy-eddy interactions, and $P_{\mathbf{k}}(t,s)$, due to eddy-mean flow and eddy-topographic interactions, are positive semi-definite in the sense of equation (19) of Bowman et al. (1993).

The equation for the diagonal response function is derived in a similar way using equation (B.2) and equation (B.4). We find (F99; FO05)

$$\left(\frac{\partial}{\partial t}+\nu_0(\mathbf{k})k^2\right)R_{\mathbf{k}}(t,t') + \int_{t'}^{t}ds(\eta_{\mathbf{k}}(t,s)+\pi_{\mathbf{k}}(t,s))R_{\mathbf{k}}(s,t') = \delta(t-t') \quad (4.11)$$

for $t \geq t'$ where the delta function implies that $R_{\mathbf{k}}(t,t)=1$. Finally, the singe-time cumulant equation takes the form

$$\left(\frac{\partial}{\partial t}+2\operatorname{Re}\nu_0(\mathbf{k})k^2\right)C_{\mathbf{k}}(t,t) = 2\operatorname{Re}\int_{t_0}^{t}ds(S_{\mathbf{k}}(t,s)+P_{\mathbf{k}}(t,s)+F_{\mathbf{k}}^0(t,s))R_{-\mathbf{k}}(t,s)$$
$$-2\operatorname{Re}\int_{t_0}^{t}ds(\eta_{\mathbf{k}}(t,s)+\pi_{\mathbf{k}}(t,s))C_{-\mathbf{k}}(t,s). \quad (4.12)$$

The CUQDIA closure, the cumulant update version of the QDIA, uses non-Gaussian and inhomogeneous cumulants, accumulated in the integrations, in the initial conditions of the periodic restarts, as described in the Appendix of FO05. For the sake of brevity these terms are not included here since they are not needed for the development of our three MIC models that follows next.

## 5. Markovian inhomogeneous closure equations

In this Section, we formulate three slightly different versions of the Markovian inhomogeneous closure (MIC) equations. We employ three commonly used versions of the FDT together with a Markovian version of the response function equation in the derivation. We begin by reformulating the single-time covariance equation (4.10) which we first write in the form

$$\frac{\partial}{\partial t}C_{\mathbf{k}}(t,t) = 2\operatorname{Re}[F_{\mathbf{k}}^{S}(t)+F_{\mathbf{k}}^{PA}(t)+F_{\mathbf{k}}^{PB}(t)+F_{\mathbf{k}}^{PC}(t)+F_{\mathbf{k}}^{PD}(t)+F_{\mathbf{k}}^{0}(t)]$$
$$-2\operatorname{Re}[N_{\mathbf{k}}^{\eta}(t)+N_{\mathbf{k}}^{\pi A}(t)+N_{\mathbf{k}}^{\pi B}(t)+N_{\mathbf{k}}^{\pi C}(t)+N_{\mathbf{k}}^{\pi D}(t)+N_{\mathbf{k}}^{0}]. \quad (5.1)$$

Here, we have first split up the $P_{\mathbf{k}}(t,s)$ and $\pi_{\mathbf{k}}(t,s)$ terms into their mean vorticity, topographic and cross terms. Simple algebra shows that the $F_{\mathbf{k}}(t)$ and $N_{\mathbf{k}}(t)$ functions have the following expressions

$$F_{\mathbf{k}}^{S}(t) = 2\sum_{\mathbf{p}}\sum_{\mathbf{q}}\delta(\mathbf{k},\mathbf{p},\mathbf{q})K(\mathbf{k},\mathbf{p},\mathbf{q})K(-\mathbf{k},-\mathbf{p},-\mathbf{q})\Delta^{S}(-\mathbf{k},-\mathbf{p},-\mathbf{q})(t),$$
$$(5.2a)$$

$$\Delta^S(-\mathbf{k},-\mathbf{p},-\mathbf{q})(t) = \int_{t_0}^{t} ds R_{-\mathbf{k}}(t,s) C_{-\mathbf{p}}(t,s) C_{-\mathbf{q}}(t,s), \tag{5.2b}$$

$$F_{\mathbf{k}}^{PA}(t) = 4 \sum_{\mathbf{p}} \sum_{\mathbf{q}} \delta(\mathbf{k},\mathbf{p},\mathbf{q}) K(\mathbf{k},\mathbf{p},\mathbf{q}) K(-\mathbf{k},-\mathbf{p},-\mathbf{q}) \Delta^{PA}(-\mathbf{k},-\mathbf{p},-\mathbf{q})(t), \tag{5.2c}$$

$$\Delta^{PA}(-\mathbf{k},-\mathbf{p},-\mathbf{q})(t) = <\zeta_{-\mathbf{q}}(t)> \int_{t_0}^{t} ds R_{-\mathbf{k}}(t,s) C_{-\mathbf{p}}(t,s) <\zeta_{\mathbf{q}}(s)>, \tag{5.2d}$$

$$F_{\mathbf{k}}^{PB}(t) = \sum_{\mathbf{p}} \sum_{\mathbf{q}} \delta(\mathbf{k},\mathbf{p},\mathbf{q}) A(\mathbf{k},\mathbf{p},\mathbf{q}) A(-\mathbf{k},-\mathbf{p},-\mathbf{q}) \Delta^{PB}(-\mathbf{k},-\mathbf{p},-\mathbf{q})(t), \tag{5.2e}$$

$$\Delta^{PB}(-\mathbf{k},-\mathbf{p},-\mathbf{q})(t) = h_{-\mathbf{q}} h_{\mathbf{q}} \int_{t_0}^{t} ds R_{-\mathbf{k}}(t,s) C_{-\mathbf{p}}(t,s), \tag{5.2f}$$

$$F_{\mathbf{k}}^{PC}(t) = 2 \sum_{\mathbf{p}} \sum_{\mathbf{q}} \delta(\mathbf{k},\mathbf{p},\mathbf{q}) K(\mathbf{k},\mathbf{p},\mathbf{q}) A(-\mathbf{k},-\mathbf{p},-\mathbf{q}) \Delta^{PC}(-\mathbf{k},-\mathbf{p},-\mathbf{q})(t), \tag{5.2g}$$

$$\Delta^{PC}(-\mathbf{k},-\mathbf{p},-\mathbf{q})(t) = <\zeta_{-\mathbf{q}}(t)> h_{\mathbf{q}} \int_{t_0}^{t} ds R_{-\mathbf{k}}(t,s) C_{-\mathbf{p}}(t,s), \tag{5.2h}$$

$$F_{\mathbf{k}}^{PD}(t) = 2 \sum_{\mathbf{p}} \sum_{\mathbf{q}} \delta(\mathbf{k},\mathbf{p},\mathbf{q}) A(\mathbf{k},\mathbf{p},\mathbf{q}) K(-\mathbf{k},-\mathbf{p},-\mathbf{q}) \Delta^{PD}(-\mathbf{k},-\mathbf{p},-\mathbf{q})(t), \tag{5.2i}$$

$$\Delta^{PD}(-\mathbf{k},-\mathbf{p},-\mathbf{q})(t) = h_{-\mathbf{q}} \int_{t_0}^{t} ds R_{-\mathbf{k}}(t,s) C_{-\mathbf{p}}(t,s) <\zeta_{\mathbf{q}}(s)>, \tag{5.2j}$$

$$F_{\mathbf{k}}^{\Theta}(t) = \int_{t_0}^{t} ds F_{\mathbf{k}}^{0}(t,s) R_{-\mathbf{k}}(t,s). \tag{5.2k}$$

Also

$$N_{\mathbf{k}}^{\eta}(t) = -4 \sum_{\mathbf{p}} \sum_{\mathbf{q}} \delta(\mathbf{k},\mathbf{p},\mathbf{q}) K(\mathbf{k},\mathbf{p},\mathbf{q}) K(-\mathbf{p},-\mathbf{q},-\mathbf{k}) \Delta^{\eta}(-\mathbf{p},-\mathbf{q},-\mathbf{k})(t), \tag{5.3a}$$

$$\Delta^{\eta}(-\mathbf{p},-\mathbf{q},-\mathbf{k})(t) = \int_{t_0}^{t} ds R_{-\mathbf{p}}(t,s) C_{-\mathbf{q}}(t,s) C_{-\mathbf{k}}(t,s) \equiv \Delta^{S}(-\mathbf{p},-\mathbf{q},-\mathbf{k})(t), \tag{5.3b}$$

$$N_{\mathbf{k}}^{\pi A}(t) = -4 \sum_{\mathbf{p}} \sum_{\mathbf{q}} \delta(\mathbf{k},\mathbf{p},\mathbf{q}) K(\mathbf{k},\mathbf{p},\mathbf{q}) K(-\mathbf{p},-\mathbf{k},-\mathbf{q}) \Delta^{\pi A}(-\mathbf{p},-\mathbf{k},-\mathbf{q})(t), \tag{5.3c}$$

$$\Delta^{\pi A}(-\mathbf{p},-\mathbf{k},-\mathbf{q})(t) = <\zeta_{-\mathbf{q}}(t)> \int_{t_0}^{t} ds R_{-\mathbf{p}}(t,s) C_{-\mathbf{k}}(t,s) <\zeta_{\mathbf{q}}(s)> \equiv \Delta^{PA}(-\mathbf{p},-\mathbf{k},-\mathbf{q})(t), \tag{5.3d}$$

$$N_{\mathbf{k}}^{\pi B}(t) = -\sum_{\mathbf{p}} \sum_{\mathbf{q}} \delta(\mathbf{k},\mathbf{p},\mathbf{q}) A(\mathbf{k},\mathbf{p},\mathbf{q}) A(-\mathbf{p},-\mathbf{k},-\mathbf{q}) \Delta^{\pi B}(-\mathbf{p},-\mathbf{k},-\mathbf{q})(t), \tag{5.3e}$$

$$\Delta^{\pi B}(-\mathbf{p},-\mathbf{k},-\mathbf{q})(t) = h_{-\mathbf{q}} h_{\mathbf{q}} \int_{t_0}^{t} ds C_{-\mathbf{k}}(t,s) R_{-\mathbf{p}}(t,s) \equiv \Delta^{PB}(-\mathbf{p},-\mathbf{k},-\mathbf{q})(t), \tag{5.3f}$$

$$N_{\mathbf{k}}^{\pi C}(t) = -2\sum_{\mathbf{p}}\sum_{\mathbf{q}} \delta(\mathbf{k},\mathbf{p},\mathbf{q}) K(\mathbf{k},\mathbf{p},\mathbf{q}) A(-\mathbf{p},-\mathbf{k},-\mathbf{q}) \Delta^{\pi C}(-\mathbf{p},-\mathbf{k},-\mathbf{q})(t), \quad (5.3g)$$

$$\Delta^{\pi C}(-\mathbf{p},-\mathbf{k},-\mathbf{q})(t) = <\zeta_{-\mathbf{q}}(t)> h_{\mathbf{q}} \int_{t_0}^{t} ds R_{-\mathbf{p}}(t,s) C_{-\mathbf{k}}(t,s) \equiv \Delta^{PC}(-\mathbf{p},-\mathbf{k},-\mathbf{q})(t), \quad (5.3h)$$

$$N_{\mathbf{k}}^{\pi D}(t) = -2\sum_{\mathbf{p}}\sum_{\mathbf{q}} \delta(\mathbf{k},\mathbf{p},\mathbf{q}) A(\mathbf{k},\mathbf{p},\mathbf{q}) K(-\mathbf{p},-\mathbf{k},-\mathbf{q}) \Delta^{\pi D}(-\mathbf{p},-\mathbf{k},-\mathbf{q})(t), \quad (5.3i)$$

$$\Delta^{\pi D}(-\mathbf{p},-\mathbf{k},-\mathbf{q})(t) = h_{-\mathbf{q}} \int_{t_0}^{t} ds R_{-\mathbf{p}}(t,s) C_{-\mathbf{k}}(t,s) <\zeta_{\mathbf{q}}(s)> \equiv \Delta^{PD}(-\mathbf{p},-\mathbf{k},-\mathbf{q})(t), \quad (5.3j)$$

$$N_{\mathbf{k}}^{\rho}(t) = \nu_0(\mathbf{k}) k^2 C_{\mathbf{k}}(t,t). \quad (5.3k)$$

The three versions of the fluctuation dissipation theorem (FDT) can be combined as follows

$$C_{\mathbf{k}}(t,t') \equiv [C_{\mathbf{k}}(t,t)]^{1-X} R_{\mathbf{k}}(t,t') [C_{\mathbf{k}}(t',t')]^{X} \quad (5.4)$$

for $t \geq t'$ and $C_{\mathbf{k}}(t,t') = C_{-\mathbf{k}}(t',t)$ for $t' > t$. Here $X = 0$ corresponds to the *current-time FDT* (Equation (A.16) of Frederiksen & Davies, 1997 and references therein) usually used in the EDQNM, $X = 1/2$ is the *correlation FDT* in equation (61) of Bowman et al. (1993) and $X = 1$ is the *prior-time FDT* (Equation (3.5) of Carnevale & Frederiksen 1983a and references therein). Bowman et al. (1993) point out that in the presence of wave phenomena it is possible for the forms with $X = 0$ and $X = 1$ to lead to unphysical results while the form with $X = 1/2$ (together with Markovian response functions with positive damping) will always be realizable. Their demonstration showing that $C_{\mathbf{k}}(t,t)$ is real and non-negative when $X = 1/2$, in their equation (65), applies equally in the QDIA inhomogeneous formalism. This follows by replacing their eddy-eddy nonlinear noise $F_{\mathbf{k}}$ by our total nonlinear noise $[S_{\mathbf{k}} + P_{\mathbf{k}}]$. This is an important point in general but whether unphysical results ensue will of course also depend on the parameter regime of the flow. For that reason we examine all three cases in this study.

With equation (5.4) substituted into equations (5.2) and (5.3) it is then possible to express the nonlinear noises and dampings in forms that involve triad relaxation functions $\Theta^X$, $\Phi^X$, and $\Psi^X$. Thus,

$$F_{\mathbf{k}}^{S}(t) = 2\sum_{\mathbf{p}}\sum_{\mathbf{q}} \delta(\mathbf{k},\mathbf{p},\mathbf{q}) K(\mathbf{k},\mathbf{p},\mathbf{q}) K(-\mathbf{k},-\mathbf{p},-\mathbf{q}) C_{-\mathbf{p}}^{1-X}(t,t) C_{-\mathbf{q}}^{X}(t,t) \Theta^{X}(-\mathbf{k},-\mathbf{p},-\mathbf{q})(t), \quad (5.5a)$$

$$\Theta^{X}(-\mathbf{k},-\mathbf{p},-\mathbf{q})(t) = \int_{t_0}^{t} ds R_{-\mathbf{k}}(t,s) R_{-\mathbf{p}}(t,s) R_{-\mathbf{q}}(t,s) C_{-\mathbf{p}}^{X}(s,s) C_{-\mathbf{q}}^{X}(s,s), \quad (5.5b)$$

$$F_{\mathbf{k}}^{PA}(t) = 4\sum_{\mathbf{p}}\sum_{\mathbf{q}} \delta(\mathbf{k},\mathbf{p},\mathbf{q}) K(\mathbf{k},\mathbf{p},\mathbf{q}) K(-\mathbf{k},-\mathbf{p},-\mathbf{q}) C_{-\mathbf{p}}^{1-X}(t,t) <\zeta_{-\mathbf{q}}(t)> \Phi^{X}(-\mathbf{k},-\mathbf{p},-\mathbf{q})(t), \quad (5.5c)$$

$$\Phi^{X}(-\mathbf{k},-\mathbf{p},-\mathbf{q})(t) = \int_{t_0}^{t} ds R_{-\mathbf{k}}(t,s) R_{-\mathbf{p}}(t,s) C_{-\mathbf{p}}^{X}(s,s) <\zeta_{\mathbf{q}}(s)>, \quad (5.5d)$$

$$F_{\mathbf{k}}^{PB}(t) = \sum_{\mathbf{p}}\sum_{\mathbf{q}} \delta(\mathbf{k},\mathbf{p},\mathbf{q}) A(\mathbf{k},\mathbf{p},\mathbf{q}) A(-\mathbf{k},-\mathbf{p},-\mathbf{q}) C_{-\mathbf{p}}^{1-X}(t,t) h_{-\mathbf{q}} h_{\mathbf{q}} \Psi^{X}(-\mathbf{k},-\mathbf{p},-\mathbf{q})(t), \quad (5.5e)$$

$$\Psi^{X}(-\mathbf{k},-\mathbf{p},-\mathbf{q})(t) = \int_{t_0}^{t} ds\, R_{-\mathbf{k}}(t,s) R_{-\mathbf{p}}(t,s) C_{-\mathbf{p}}^{X}(s,s), \quad (5.5f)$$

$$F_{\mathbf{k}}^{PC}(t) = 2\sum_{\mathbf{p}}\sum_{\mathbf{q}} \delta(\mathbf{k},\mathbf{p},\mathbf{q}) K(\mathbf{k},\mathbf{p},\mathbf{q}) A(-\mathbf{k},-\mathbf{p},-\mathbf{q}) C_{-\mathbf{p}}^{1-X}(t,t) <\zeta_{-\mathbf{q}}(t)> h_{\mathbf{q}} \Psi^{X}(-\mathbf{k},-\mathbf{p},-\mathbf{q})(t) \quad (5.5g)$$

$$F_{\mathbf{k}}^{PD}(t) = 2\sum_{\mathbf{p}}\sum_{\mathbf{q}} \delta(\mathbf{k},\mathbf{p},\mathbf{q}) A(\mathbf{k},\mathbf{p},\mathbf{q}) K(-\mathbf{k},-\mathbf{p},-\mathbf{q}) C_{-\mathbf{p}}^{1-X}(t,t) h_{-\mathbf{q}} \Phi^{X}(-\mathbf{k},-\mathbf{p},-\mathbf{q})(t), \quad (5.5h)$$

$$F_{\mathbf{k}}^{0}(t) = \int_{t_0}^{t} ds\, F_{\mathbf{k}}^{0}(t,s) R_{-\mathbf{k}}(t,s). \quad (5.5i)$$

Also
$$N_{\mathbf{k}}^{\eta}(t) = D_{\mathbf{k}}^{\eta}(t) C_{\mathbf{k}}(t,t) \quad (5.6a)$$

where
$$D_{\mathbf{k}}^{\eta}(t) = -4\sum_{\mathbf{p}}\sum_{\mathbf{q}} \delta(\mathbf{k},\mathbf{p},\mathbf{q}) K(\mathbf{k},\mathbf{p},\mathbf{q}) K(-\mathbf{p},-\mathbf{q},-\mathbf{k}) C_{-\mathbf{q}}^{1-X}(t,t) C_{-\mathbf{k}}^{-X}(t,t) \Theta^{X}(-\mathbf{p},-\mathbf{q},-\mathbf{k})(t), \quad (5.6b)$$

$$N_{\mathbf{k}}^{TA}(t) = D_{\mathbf{k}}^{TA}(t) C_{\mathbf{k}}(t,t) \quad (5.6c)$$

where
$$D_{\mathbf{k}}^{TA}(t) = -4\sum_{\mathbf{p}}\sum_{\mathbf{q}} \delta(\mathbf{k},\mathbf{p},\mathbf{q}) K(\mathbf{k},\mathbf{p},\mathbf{q}) K(-\mathbf{p},-\mathbf{k},-\mathbf{q}) C_{-\mathbf{k}}^{-X}(t,t) <\zeta_{-\mathbf{q}}(t)> \Phi^{X}(-\mathbf{p},-\mathbf{k},-\mathbf{q})(t), \quad (5.6d)$$

$$N_{\mathbf{k}}^{TB}(t) = D_{\mathbf{k}}^{TB}(t) C_{\mathbf{k}}(t,t) \quad (5.6e)$$

where
$$D_{\mathbf{k}}^{TB}(t) = -\sum_{\mathbf{p}}\sum_{\mathbf{q}} \delta(\mathbf{k},\mathbf{p},\mathbf{q}) A(\mathbf{k},\mathbf{p},\mathbf{q}) A(-\mathbf{p},-\mathbf{k},-\mathbf{q}) h_{-\mathbf{q}} h_{\mathbf{q}} C_{-\mathbf{k}}^{-X}(t,t) \Psi^{X}(-\mathbf{p},-\mathbf{k},-\mathbf{q})(t), \quad (5.6f)$$

$$N_{\mathbf{k}}^{TC}(t) = D_{\mathbf{k}}^{TC}(t) C_{\mathbf{k}}(t,t) \quad (5.6g)$$

where
$$D_{\mathbf{k}}^{TC}(t) = -2\sum_{\mathbf{p}}\sum_{\mathbf{q}} \delta(\mathbf{k},\mathbf{p},\mathbf{q}) K(\mathbf{k},\mathbf{p},\mathbf{q}) A(-\mathbf{p},-\mathbf{k},-\mathbf{q}) <\zeta_{-\mathbf{q}}(t)> h_{\mathbf{q}} C_{-\mathbf{k}}^{-X}(t,t) \Psi^{X}(-\mathbf{p},-\mathbf{k},-\mathbf{q})(t), \quad (5.6h)$$

$$N_{\mathbf{k}}^{TD}(t) = D_{\mathbf{k}}^{TD}(t) C_{\mathbf{k}}(t,t) \quad (5.6i)$$

where
$$D_{\mathbf{k}}^{TD}(t) = -2\sum_{\mathbf{p}}\sum_{\mathbf{q}} \delta(\mathbf{k},\mathbf{p},\mathbf{q}) A(\mathbf{k},\mathbf{p},\mathbf{q}) K(-\mathbf{p},-\mathbf{k},-\mathbf{q}) h_{-\mathbf{q}} C_{-\mathbf{k}}^{-X}(t,t) \Phi^{X}(-\mathbf{p},-\mathbf{k},-\mathbf{q})(t), \quad (5.6j)$$

$$N_{\mathbf{k}}^{0}(t) = D_{\mathbf{k}}^{0}(t) C_{\mathbf{k}}(t,t) \quad (5.6k)$$

where
$$D_{\mathbf{k}}^0(t) = \nu_0(\mathbf{k})k^2. \qquad (5.6l)$$

Note that $C_{\mathbf{k}}(t,t) = C_{-\mathbf{k}}(t,t)$ since $C_{\mathbf{k}}(t,t)$ is real. It convenient then to define

$$F_{\mathbf{k}}^1(t) = F_{\mathbf{k}}^S(t); \; F_{\mathbf{k}}^2(t) = F_{\mathbf{k}}^{PA}(t); \; F_{\mathbf{k}}^3(t) = F_{\mathbf{k}}^{PA}(t); \; F_{\mathbf{k}}^4(t) = F_{\mathbf{k}}^{PC}(t); \; F_{\mathbf{k}}^5(t) = F_{\mathbf{k}}^{PD}(t); \qquad (5.7a)$$
$$D_{\mathbf{k}}^1(t) = D_{\mathbf{k}}^\eta(t); \; D_{\mathbf{k}}^2(t) = D_{\mathbf{k}}^{TA}(t); \; D_{\mathbf{k}}^3(t) = D_{\mathbf{k}}^{TB}(t); \; D_{\mathbf{k}}^4(t) = D_{\mathbf{k}}^{TC}(t); \; D_{\mathbf{k}}^5(t) = D_{\mathbf{k}}^{TD}(t).$$

From equations (4.12) and (5.1) we see that
$$\sum_{j=0}^{5} D_{\mathbf{k}}^j(t) = \nu_0(\mathbf{k})k^2 + \int_{t_0}^{t} ds\{(\eta_{\mathbf{k}}(t,s) + \pi_{\mathbf{k}}(t,s))C_{-\mathbf{k}}(t,s)[C_{\mathbf{k}}(t,t)]^{-1}\}. \qquad (5.7b)$$

Thus
$$\frac{\partial}{\partial t} C_{\mathbf{k}}(t,t) = 2\text{Re}[\sum_{j=0}^{5} \{F_{\mathbf{k}}^j(t) - D_{\mathbf{k}}^j(t)C_{\mathbf{k}}(t,t)\}]. \qquad (5.8)$$

Now to complete the Markovianization, the response function equation (4.11) must also be replaced by
$$\frac{\partial}{\partial t} R_{\mathbf{k}}(t,t') + \sum_{j=0}^{5} D_{\mathbf{k}}^j(t)R_{\mathbf{k}}(t,t') = \delta(t-t'). \qquad (5.9)$$

From equation (5.7b) we see that equation (5.9) is equivalent to
$$\frac{\partial}{\partial t} R_{\mathbf{k}}(t,t') + \left[\nu_0(\mathbf{k})k^2 + \int_{t_0}^{t} ds\{(\eta_{\mathbf{k}}(t,s) + \pi_{\mathbf{k}}(t,s))C_{-\mathbf{k}}(t,s)[C_{\mathbf{k}}(t,t)]^{-1}\}\right]R_{\mathbf{k}}(t,t')$$
$$= \delta(t-t'). \qquad (5.10)$$

The advantage of the forms (5.8) and (5.9) is however that the integral terms can be replaced by manifestly Markovian equations for the triad relaxation times that also involve the damping terms $D_{\mathbf{k}}^j(t)$. This can be seen as follows. The integral expression for the relaxation time
$$\Theta^X(\mathbf{k},\mathbf{p},\mathbf{q})(t) = \int_{t_0}^{t} ds R_{\mathbf{k}}(t,s)R_{\mathbf{p}}(t,s)R_{\mathbf{q}}(t,s)C_{\mathbf{p}}^X(s,s)C_{\mathbf{q}}^X(s,s), \qquad (5.11a)$$

is equivalent to the differential equation
$$\frac{\partial}{\partial t} \Theta^X(\mathbf{k},\mathbf{p},\mathbf{q})(t) + \sum_{j=0}^{5} [D_{\mathbf{k}}^j(t) + D_{\mathbf{p}}^j(t) + D_{\mathbf{q}}^j(t)]\Theta^X(\mathbf{k},\mathbf{p},\mathbf{q})(t) = C_{\mathbf{p}}^X(t,t)C_{\mathbf{q}}^X(t,t) \qquad (5.11b)$$

with $\Theta^X(\mathbf{k},\mathbf{p},\mathbf{q})(0) = 0$. Also
$$\Phi^X(\mathbf{k},\mathbf{p},\mathbf{q})(t) = \int_{t_0}^{t} ds R_{\mathbf{k}}(t,s)R_{\mathbf{p}}(t,s)C_{\mathbf{p}}^X(s,s)<\zeta_{-\mathbf{q}}(s)> \qquad (5.11c)$$

is equivalent to

$$\frac{\partial}{\partial t}\Phi^X(\mathbf{k},\mathbf{p},\mathbf{q})(t)+\sum_{j=0}^{5}[D_{\mathbf{k}}^j(t)+D_{\mathbf{p}}^j(t)]\Phi^X(\mathbf{k},\mathbf{p},\mathbf{q})(t)=C_{\mathbf{p}}^X(t,t)<\zeta_{-\mathbf{q}}(t)> \quad (5.11d)$$

with $\Phi^X(\mathbf{k},\mathbf{p},\mathbf{q})(0)=0$. Finally,

$$\Psi^X(\mathbf{k},\mathbf{p},\mathbf{q})(t)=\int_{t_0}^{t}ds R_{\mathbf{k}}(t,s)R_{\mathbf{p}}(t,s)C_{\mathbf{p}}^X(s,s) \quad (5.11e)$$

is equivalent to

$$\frac{\partial}{\partial t}\Psi^X(\mathbf{k},\mathbf{p},\mathbf{q})(t)+\sum_{j=0}^{5}[D_{\mathbf{k}}^j(t)+D_{\mathbf{p}}^j(t)]\Psi^X(\mathbf{k},\mathbf{p},\mathbf{q})(t)=C_{\mathbf{p}}^X(t,t) \quad (5.11f)$$

with $\Psi^X(\mathbf{k},\mathbf{p},\mathbf{q})(0)=0$.

In order to close the Markovian equations for the single-time diagonal cumulant and response function, with auxiliary equations for the triad relaxation times, we need to formulate a Markov version of the mean field equation. From equation (4.7) we have

$$\left(\frac{\partial}{\partial t}+\nu_0(\mathbf{k})k^2\right)<\zeta_{\mathbf{k}}>=\sum_{\mathbf{p}}\sum_{\mathbf{q}}\delta(\mathbf{k},\mathbf{p},\mathbf{q})[K(\mathbf{k},\mathbf{p},\mathbf{q})<\zeta_{-\mathbf{p}}><\zeta_{-\mathbf{q}}>$$
$$+A(\mathbf{k},\mathbf{p},\mathbf{q})<\zeta_{-\mathbf{p}}>h_{-\mathbf{q}}]-N_{\mathbf{k}}^M(t)+f_{\mathbf{k}}^h(t)+\bar{f}_{\mathbf{k}}^0(t). \quad (5.12)$$

Here, the nonlinear damping and eddy-topographic force are given by

$$N_{\mathbf{k}}^M(t)=-4\sum_{\mathbf{p}}\sum_{\mathbf{q}}\delta(\mathbf{k},\mathbf{p},\mathbf{q})K(\mathbf{k},\mathbf{p},\mathbf{q})K(-\mathbf{p},-\mathbf{q},-\mathbf{k})$$
$$\times\int_{t_0}^{t}ds R_{-\mathbf{p}}(t,s)C_{-\mathbf{q}}(t,s)<\zeta_{\mathbf{k}}(s)>, \quad (5.13a)$$

$$f_{\mathbf{k}}^h(t)=2\sum_{\mathbf{p}}\sum_{\mathbf{q}}\delta(\mathbf{k},\mathbf{p},\mathbf{q})K(\mathbf{k},\mathbf{p},\mathbf{q})A(-\mathbf{p},-\mathbf{q},-\mathbf{k})$$
$$\times h_{\mathbf{k}}\int_{t_0}^{t}ds R_{-\mathbf{p}}(t,s)C_{-\mathbf{q}}(t,s). \quad (5.13b)$$

Thus

$$N_{\mathbf{k}}^M(t)=-4\sum_{\mathbf{p}}\sum_{\mathbf{q}}\delta(\mathbf{k},\mathbf{p},\mathbf{q})K(\mathbf{k},\mathbf{p},\mathbf{q})K(-\mathbf{p},-\mathbf{q},-\mathbf{k})\Delta^M(-\mathbf{p},-\mathbf{q},-\mathbf{k})(t), \quad (5.14a)$$

$$\Delta^M(-\mathbf{p},-\mathbf{q},-\mathbf{k})(t)=\int_{t_0}^{t}ds R_{-\mathbf{p}}(t,s)C_{-\mathbf{q}}(t,s)<\zeta_{\mathbf{k}}(s)>, \quad (5.14b)$$

and

$$f_{\mathbf{k}}^h(t)=2\sum_{\mathbf{p}}\sum_{\mathbf{q}}\delta(\mathbf{k},\mathbf{p},\mathbf{q})K(\mathbf{k},\mathbf{p},\mathbf{q})A(-\mathbf{p},-\mathbf{q},-\mathbf{k})\Delta^h(-\mathbf{p},-\mathbf{q},-\mathbf{k})(t), \quad (5.14c)$$

$$\Delta^h(-\mathbf{p},-\mathbf{q},-\mathbf{k})(t) = h_\mathbf{k} \int_{t_0}^{t} ds R_{-\mathbf{p}}(t,s) C_{-\mathbf{q}}(t,s). \tag{5.14d}$$

Next, we again use the fluctuation dissipation theorem (5.4) for $t \geq t'$. Then we have

$$N_\mathbf{k}^M(t) = D_\mathbf{k}^M(t) <\zeta_\mathbf{k}(t)> \tag{5.15a}$$

where $D_\mathbf{k}^M(t) = -4 \sum_\mathbf{p} \sum_\mathbf{q} \delta(\mathbf{k},\mathbf{p},\mathbf{q}) K(\mathbf{k},\mathbf{p},\mathbf{q}) K(-\mathbf{p},-\mathbf{q},-\mathbf{k}) C_{-\mathbf{q}}^{1-X}(t,t) <\zeta_\mathbf{k}(t)>^{-1} \Phi^X(-\mathbf{p},-\mathbf{q},-\mathbf{k})(t),$ (5.15b)

and $f_\mathbf{k}^h(t) = 2 \sum_\mathbf{p} \sum_\mathbf{q} \delta(\mathbf{k},\mathbf{p},\mathbf{q}) K(\mathbf{k},\mathbf{p},\mathbf{q}) A(-\mathbf{p},-\mathbf{q},-\mathbf{k}) C_{-\mathbf{q}}^{1-X}(t,t) h_\mathbf{k} \Psi^X(-\mathbf{p},-\mathbf{q},-\mathbf{k})(t).$ (5.15c)

Here, the expressions for $\Phi^X$ and $\Psi^X$ are the same as in equation (5.5). Similarly, we have the manifestly Markovian equations for the triad relaxation times associated with the mean field as in equation (5.11).

Now, equation (5.12) can be written in the form

$$\frac{\partial}{\partial t} <\zeta_\mathbf{k}> = \sum_\mathbf{p} \sum_\mathbf{q} \delta(\mathbf{k},\mathbf{p},\mathbf{q})[K(\mathbf{k},\mathbf{p},\mathbf{q}) <\zeta_{-\mathbf{p}}><\zeta_{-\mathbf{q}}> \\ + A(\mathbf{k},\mathbf{p},\mathbf{q}) <\zeta_{-\mathbf{p}}> h_{-\mathbf{q}}] - [D_\mathbf{k}^0 + D_\mathbf{k}^M(t)] <\zeta_\mathbf{k}(t)> + f_\mathbf{k}^h(t) + \bar{f}_\mathbf{k}^0. \tag{5.16}$$

This then closes the equations for the three versions of the Markovian inhomogeneous closure with the unique triad relaxation times being $\Theta^X(\mathbf{k},\mathbf{p},\mathbf{q})(t), \Phi^X(\mathbf{k},\mathbf{p},\mathbf{q})(t)$ and $\Psi^X(\mathbf{k},\mathbf{p},\mathbf{q})(t)$. Of course, rather than performing the integrals over time the partial differential equations for the triad relaxation times must be solved. The relative efficiency of the methods will depend on how long the integration needs to be done before a cumulant update restart can be performed (OF04; FO05) and the dimensionality of the problem. However, the Markovian inhomogeneous closure also offers the possibility of analytic forms of the triad relaxation times, such as is typically used in EDQNM calculations, and this of course would speed up the calculations enormously.

## 6. Performance of MIC models for turbulent flow and Rossby waves over topography

We test the performance of the three versions of the MIC model, described in Section 5, against two variants of the non-Markovian QDIA and against a large ensemble of direct numerical simulations (DNS) for the case of inhomogeneous turbulent flows over an isolated topographic feature. The numerical simulation and closure calculation setup is similar to that described in FO05. We consider the turbulent interaction of an initial two-dimensional eastward mean flow $U$ impinging on isolated topography on a $\beta$– plane. As noted in FO05, where many references are given, this is an iconic problem in mean flow-topographic interaction going back to the work of Kasahara (1966). This is a far from equilibrium problem that is a severe test of the closures with the Rossby wavetrains rapidly spun up in the presence or turbulence. The topography is a circular conical mountain centered at 30°N, 180°W, with a height of 2.5 km and a diameter of 45° latitude (Figure 1 of FO05) and is also similar to that used in the linear and nonlinear study on the sphere by Frederiksen (1982).

The focus here is on a case similar to case 1 of FO05 where the initial mean flow $U$ is eastward at $7.5 \, \text{ms}^{-1}$ and the $\beta$-effect is $1.15 \times 10^{-11} \, \text{m}^{-1}\text{s}^{-1}$. Throughout we use a length scale of one half the earth's radius, $a_e/2$, and a time scale of the inverse of the earth's rotation rate, $\Omega^{-1}$, which means the non-dimensional values are $U = 0.0325$, $\beta = \frac{1}{2}$ and $k_0^2 = \frac{1}{2}$, and the $\beta$-effect is typical of that at 60° latitude. The viscosity is $2.5 \times 10^4 \, \text{m}^2\text{s}^{-1}$ or $\hat{\nu} = 3.378 \times 10^{-5}$ in non-dimensional units and $f^0 = 0 = \alpha_U$. Other details of the initial setup are specified in Table 1. As well as the large-scale flow $U$ the initial conditions include a small amplitude mean flow, which is localized over the isolated topography, and an isotropic transient spectrum, both of which are specified in Table 1. The calculations are started from this mean field to which is added Gaussian isotropic perturbations with the spectrum given in Table 1. The calculations use a circular C16 resolution, with $k \leq 16$, which is adequate for studying Rossby wave dispersion in a turbulent environment.

Some care is needed in setting up the 1800 initial conditions for the ensemble of DNS fields that the closure calculations are compared with, as noted by FO05. First a field is constructed as a Gaussian sample with zero mean and unit variance. Then from this field further initial members are obtained by moving its origin by a given increment in the $x$-direction and then in the $y$-direction. A total of $n$ realizations are obtained by successively moving the initial field by $2\pi/n$ in the $x$-direction. By subsequently moving each of these $n$ realizations by $2\pi/n$ in the $y$-direction we obtain a total of $n^2$ realizations and also taking the negative values of each gives $2n^2$ realizations. The process can be repeated until the required number of realizations are arrived at and the fields then scaled as required and added to the mean vorticity in Table 1. The method ensures that the DNS covariance matrix is nearly isotropic for sufficiently large $n$.

The time evolutions with the DNS, with two variants of the non-Markovian QDIA and with the three versions of the MIC are carried out with a time step of $1/30$ day (non-dimensional $\Delta t = 0.21$) for a total of 10 days each. The time stepping employs a predictor-corrector method and the integrals are performed using the trapezoidal rule (Frederiksen et al. 1994; OF04; FO05). The CUQDIA calculations use a restart every day as in FO05 but we also integrate the non-Markovian QDIA equations with the full 10 day integrals evaluated without restart. As noted, the CUQDIA uses the calculated off-diagonal elements of the two-point and three-point cumulants in the new initial conditions of the periodic restart procedure. This makes the code more efficient at the expense of a judicious choice of the restart time. However, here we have also performed the much larger computational task with the full non-Markovian QDIA in order to make a very clean comparison with the Markovian closures.

Figure 1 shows the Rossby wavetrains in the zonally asymmetric component of the mean streamfunction on day 10 for the non-Markovian CUQDIA and QDIA closures, for the three variants of the MIC model and for the ensemble of DNS. The wave patterns shows the typical Rossby wave dispersion found in both linearized models and in nonlinear simulations (FO05 and references therein). The results in all the six panels are extremely similar. The numbers in brackets above each Figure part are the pattern correlations between each of the closure results and those from the ensemble of DNS. The agreement between the DNS and the three Markovian inhomogeneous closures and the non-Markovian QDIA and CUQDIA closures is excellent with pattern correlations greater than 0.9998 in all cases.

Next we examine the evolution of the transient and mean statistics. We track the evolution of the statistics through the band-averaged mean and transient kinetic energy spectra that are defined as

$$E^T(k_i,t) = \tfrac{1}{2}\sum_{\mathbf{k}\in\mathcal{S}} C_\mathbf{k}(t,t) k^{-2}, \quad (6.1a)$$

$$E^M(k_i,t) = \tfrac{1}{2}\sum_{\mathbf{k}\in\mathcal{S}} \langle\zeta_\mathbf{k}(t)\rangle\langle\zeta_{-\mathbf{k}}(t)\rangle k^{-2}. \quad (6.1b)$$

The set $\mathcal{S}$ is defined by

$$\mathcal{S} = [\mathbf{k} \,|\, k_i = \mathrm{Int}(k+\tfrac{1}{2})]. \quad (6.2)$$

Here, the average is taken over all **k** that lie within a band of unit width at a radius $k_i$ and the energy of the large scale flow is plotted at zero wavenumber.

Figure 2 shows the initial and 10 day evolved mean and transient kinetic energy spectra for the non-Markovian CUQDIA and QDIA closures and each of the three variants of the MIC model, and spectra for the ensemble of DNS are also depicted in each panel for comparison. There are only slight differences in the six sets of results on day 10. The CUQDIA kinetic spectra, both mean and transient, are virtually identical to those of the ensemble of DNS as are the QDIA spectra with just the slightest increase in the transients between wavenumbers 2 to 6. The realizable MIC that employs the correlation FDT ($X=1/2$) and the MIC that uses the prior-time FDT ($X=1$) again perform exceptionally. For both, the evolved mean kinetic energy spectra are virtually indistinguishable from the ensemble of DNS as are the transient spectra but with some slight increases between wavenumbers 2 to 6. This range of wavenumbers is of course where the changes in the mean kinetic energy are largest as the Rossby waves amplify and it is also where there is some increase in the turbulent kinetic energy. The MIC that uses the current-time FDT ($X=0$) again has evolved mean and transient kinetic energy spectra in close agreement with the ensemble of DNS but with somewhat weaker transients between wavenumbers 2 and 6 than the DNS and other closures. The method of Markovianization for this MIC (with $X=0$) is of course closest to that employed for the EDQNM where the current-time FDT in equation (1.2) is also used. Interestingly the large amplitude Rossby waves between wavenumbers 2 and 6 will tend to reduce the kinetic energy transfer into those wavenumbers as is the case for the EDQNM (Frederiksen et al. 1996 and references therein).

Altogether, the performance of the three versions of the MIC model, and of the non-Markovian CUQDIA and QDIA closures, is quite remarkable in this parameter regime of large-scale Rossby wave dispersion over topography in a turbulent environment. For higher resolution and higher Reynolds numbers we expect that the Markovian inhomogeneous closures, like the non-Markovian DIA (Frederiksen & Davies 2004) and QDIA (OF04), will need to incorporate a regularization, or empirical vertex renormalization, in order to yield the correct small scale spectra. The regularized Markovian inhomogeneous closures are documented in Appendix C.

The results here show the potential of the Markovian inhomogeneous closures to reproduce the mean and transient statistics of ensembles of DNS in a strongly inhomogeneous context. If broadly universal and analytic expressions for the triad relaxation times, or the underlying response functions, can be established in the presence of waves and inhomogeneities, then very efficient closures, like the EDQNM, can be developed to tackle the general problem of inhomogeneous turbulent flows.

## 7. Discussion and conclusions

We have formulated manifestly Markovian inhomogeneous closure models for turbulent flows and Rossby waves over topography on a generalized $\beta$-plane. These have been derived as modifications of the non-Markovian QDIA closure (F99; OF04; FO05) in which the diagonal two-time covariance is replaced by one of three versions of the fluctuation-dissipation theorem (FDT) and the diagonal response function equation is modified to a form that is Markovian. The three different Markovian inhomogeneous closures employ respectively the current-time FDT (quasi-stationary), the prior-time FDT (non-stationary) and the correlation FDT. Bowman et al. (1993) pointed out that the correlation FDT for homogeneous turbulence, in the presence of waves, together with positive damping in Markovian response functions, resulted in a realizable Markovian closure that guaranteed real and non-negative cumulants $C_\mathbf{k}(t,t)$. We have noted that the cumulants $C_\mathbf{k}(t,t)$ are again realizable in the more complex inhomogeneous MIC formalism with the correlation FDT, but, as in the homogeneous case, not necessarily with the current-time FDT or the prior-time FDT. Of course whether unphysical results arise will depend on the parameter regimes of the flows and in this article we have documented results for each of the three Markovian closures in the same regimes.

The Markovian inhomogeneous closures differ from the non-Markovian QDIA closure in that the response function has been modified to a form that is Markovian and the time history integrals have also been modified by the FDTs in such a way that their information can be characterized by three triad relaxation functions (for each variant) that satisfy auxiliary Markovian tendency equations. Thus the Markovian inhomogeneous closures contain much the same information as the non-Markovian QDIA but the time history integrals can be replaced by differential equations that become relatively more efficient for long integrations. As well there is the prospect of developing analytical forms of the triad relaxation functions, or underpinning response functions, as is the case for isotropic EDQNM closures, that would increase the computational efficiency enormously.

The performance of the three Markovian inhomogeneous closures has been compared with results from an 1800 member ensemble of direct numerical simulations (DNS) and with the non-Markovian

QDIA and CUQDIA closures for turbulent flow over isolated topography on a generalized $\beta$– plane. For each of these calculations, the initial flow consists of an eastward mean flow $U$, together with smaller amplitude mean and transient 'small scales'. The impact of the mean flow on the conical mountain topography results in the rapid generation of large amplitude Rossby waves in a turbulent environment in 10 day integrations. The calculations are performed at a C16 resolution with wavenumbers $k \leq 16$ which is sufficient for this large scale process. This is a far from equilibrium process (Frederiksen 1982; FO05) which severely tests the closures. The performance of each of the three Markovian inhomogeneous closures, and the non-Markovian QDIA integrated without a restart over the 10 days, is excellent, like the CUQDIA closure calculations with periodic 1 day restarts (FO05). In all cases the pattern correlations of the day 10 mean Rossby wave streamfunction for the closures with the ensemble result for 1800 DNS are greater than 0.9998. Over the 10 days, there is significant evolution of the mean and transient energy spectra particularly between wavenumbers 2 to 6 where the Rossby waves amplify by orders of magnitude in the mean; in this range there are some slight differences in the transient kinetic energy between the MIC calculations and the ensemble of 1800 DNS.

We have also formulated a regularized version of the inhomogeneous closures, along the lines of the regularized DIA closure of Frederiksen & Davies (2004) and the regularized QDIA closure of OF04 which are needed for high resolution and high Reynolds number calculations. The regularized closures have a wavenumber cut-off parameter $\alpha$ which localizes the interactions, and corresponds to an empirical vertex renormalization (equation (C.1)). It has been found that the value of $\alpha$ that is consistent with inertial range behavior in the DIA (Frederiksen & Davies 2004) and QDIA (OF04) is essentially universal.

In future studies we aim to experiment with different analytical expressions of the triad relaxation functions, or underpinning Markovian response functions, to see if it is possible to establish Markovian closures for the inhomogeneous problem in the presence of waves that have similar computational efficiency to the EDQNM closure for isotropic turbulence. We also aim to examine the performance of regularized versions of the Markovian closures at high resolution and Reynolds numbers. We further seek to generalize the LET closure of McComb (1974, 1990, 2014) to inhomogeneous turbulent flows and compare the performance of non-Markovian and Markovian variants.

## 8. Acknowledgments

TJO is supported by the CSIRO Decadal Forecasting project (https://research.csiro.au/dfp).

## 9. Appendix A: Interaction coefficients

As discussed in FO05, when the spectral equations for the small scales and the large scale flow are combined into a single equation as in equation (3.3), the required interaction coefficients are

$$A(\mathbf{k},\mathbf{p},\mathbf{q}) = -\gamma(p_x\hat{q}_y - \hat{p}_y q_x)/p^2, \tag{A.1a}$$

$$K(\mathbf{k},\mathbf{p},\mathbf{q}) = \tfrac{1}{2}[A(\mathbf{k},\mathbf{p},\mathbf{q}) + A(\mathbf{k},\mathbf{q},\mathbf{p})] = \tfrac{1}{2}\gamma[p_x\hat{q}_y - \hat{p}_y q_x](p^2 - q^2)/p^2 q^2 \tag{A.1b}$$

and
$$\delta(\mathbf{k},\mathbf{p},\mathbf{q}) = \begin{cases} 1 & \text{if } \mathbf{k}+\mathbf{p}+\mathbf{q}=0 \\ 0 & \text{otherwise}. \end{cases} \tag{A.1c}$$

Here, the zero wave vector representing the large scale flow is included by defining $\gamma$, $\hat{q}_y$ and $\hat{p}_y$ as follows

$$\gamma = \begin{cases} -\tfrac{1}{2}k_0 & \text{if } \mathbf{k}=\mathbf{0}, \\ k_0 & \text{if } \mathbf{q}=\mathbf{0} \text{ or } \mathbf{p}=\mathbf{0}, \\ 1 & \text{otherwise}, \end{cases} \tag{A.2a}$$

$$\hat{p}_y = \begin{cases} 1 & \text{if } \mathbf{k}=\mathbf{0}, \text{ or } \mathbf{p}=\mathbf{0}, \text{ or } \mathbf{q}=\mathbf{0} \\ p_y & \text{otherwise}, \end{cases} \tag{A.2b}$$

$$\hat{q}_y = \begin{cases} 1 & \text{if } \mathbf{k}=\mathbf{0}, \text{ or } \mathbf{p}=\mathbf{0}, \text{ or } \mathbf{q}=\mathbf{0} \\ q_y & \text{otherwise}. \end{cases} \tag{A.2c}$$

For all values of $\mathbf{k}, \mathbf{p}$ and $\mathbf{q}$, including the zero vectors, the interaction coefficients satisfy the relationships:

$$A(-\mathbf{k},-\mathbf{p},-\mathbf{q}) = A(\mathbf{k},\mathbf{p},\mathbf{q}), \tag{A.3a}$$

$$K(-\mathbf{k},-\mathbf{p},-\mathbf{q}) = K(\mathbf{k},\mathbf{p},\mathbf{q}), \tag{A.3b}$$

and
$$K(\mathbf{k},\mathbf{p},\mathbf{q}) + K(\mathbf{p},\mathbf{q},\mathbf{k}) + K(\mathbf{q},\mathbf{k},\mathbf{p}) = 0. \tag{A.3c}$$

## 10. Appendix B: QDIA two- and three-point cumulant and response function terms

Here we summarize expressions and relations for the two-point and three-point cumulants and response functions that are needed to close the QDIA equations. The first two relationships determine the QDIA off-diagonal elements of the two-point cumulant and response function through

$$C_{\mathbf{k},-\mathbf{l}}(t,t') = \int_{t_0}^{t} ds R_{\mathbf{k}}(t,s) C_{\mathbf{l}}(s,t')[A(\mathbf{k},-\mathbf{l},\mathbf{l}-\mathbf{k})h_{(\mathbf{k}-\mathbf{l})} + 2K(\mathbf{k},-\mathbf{l},\mathbf{l}-\mathbf{k})<\zeta_{(\mathbf{k}-\mathbf{l})}(s)>]$$
$$+ \int_{t_0}^{t'} ds R_{-\mathbf{l}}(t',s) C_{\mathbf{k}}(t,s)[A(-\mathbf{l},\mathbf{k},\mathbf{l}-\mathbf{k})h_{(\mathbf{k}-\mathbf{l})} + 2K(-\mathbf{l},\mathbf{k},\mathbf{l}-\mathbf{k})<\zeta_{(\mathbf{k}-\mathbf{l})}(s)>] \tag{B.1}$$

and

$$R_{\mathbf{k},\mathbf{l}}(t,t') = \int_{t'}^{t} ds R_{\mathbf{k}}(t,s) R_{\mathbf{l}}(s,t')[A(\mathbf{k},-\mathbf{l},\mathbf{l}-\mathbf{k})h_{(\mathbf{k}-\mathbf{l})} + 2K(\mathbf{k},-\mathbf{l},\mathbf{l}-\mathbf{k})<\zeta_{(\mathbf{k}-\mathbf{l})}(s)>]. \tag{B.2}$$

These relationships between the first order renormalized off-diagonal elements of the two-point cumulant, or covariance, and response function and the diagonal components and mean field and topography were derived in F99. They apply for the case of homogeneous initial conditions. For inhomogeneous initial conditions an extra term appears in the two-point cumulant as given in equation (5.4) of FO05.

We shall also need expressions for the three-point cumulant and between the response function and perturbation field which for the QDIA are

$$\langle \tilde{\zeta}_{-\mathbf{l}}(t)\tilde{\zeta}_{(\mathbf{l}-\mathbf{k})}(t)\tilde{\zeta}_{\mathbf{k}}(t')\rangle = 2\int_{t_0}^{t} ds K(\mathbf{k},-\mathbf{l},\mathbf{l}-\mathbf{k})C_{-\mathbf{l}}(t,s)C_{(\mathbf{l}-\mathbf{k})}(t,s)R_{\mathbf{k}}(t',s)$$

$$+ 2\int_{t_0}^{t} ds K(-\mathbf{l},\mathbf{l}-\mathbf{k},\mathbf{k})R_{-\mathbf{l}}(t,s)C_{(\mathbf{l}-\mathbf{k})}(t,s)C_{\mathbf{k}}(t',s) \quad (B.3)$$

$$+ 2\int_{t_0}^{t} ds K(\mathbf{k},-\mathbf{l},\mathbf{l}-\mathbf{k})C_{-\mathbf{l}}(t,s)R_{(\mathbf{l}-\mathbf{k})}(t,s)C_{\mathbf{k}}(t',s)$$

and

$$\langle \tilde{R}_{(\mathbf{l}-\mathbf{k})}(t,t')\tilde{\zeta}_{-\mathbf{l}}(t)\rangle = 2\int_{t'}^{t} ds K(\mathbf{l}-\mathbf{k},-\mathbf{l},\mathbf{k})C_{-\mathbf{l}}(t,s)R_{(\mathbf{l}-\mathbf{k})}(t,s)R_{\mathbf{k}}(s,t'). \quad (B.4)$$

A simple derivation of these relationships that apply to the DIA, and to the QDIA, appears in Frederiksen (2003) for the case of Gaussian initial conditions. For non-Gaussian initial conditions the initial three-point function also appears as shown in equation (5.8) of FO05. A complete derivation of the two-point and three-point terms needed in the cumulant update QDIA closure is also presented by O'Kane and Frederiksen (2010).

## 11. Appendix C: Regularization of Markovian inhomogeneous closures

Here, we summarize the regularization of the Markovian inhomogeneous closures that will be needed to simulate the small scales of high Reynolds number turbulence just as is needed for the non-Markovian DIA (Frederiksen & Davies 2004) and QDIA (OF04) closures. The regularization involves modifying the interaction coefficients – vertices – to localize the interactions. The reasons for this is discussed in some detail by Frederiksen & Davies (2004 and references therein). Thus the interaction coefficients are modified to

$$\breve{A}(\mathbf{k},\mathbf{p},\mathbf{q}) = \theta(p-k/\alpha)\theta(q-k/\alpha)A(\mathbf{k},\mathbf{p},\mathbf{q}), \quad (C.1a)$$

$$\breve{K}(\mathbf{k},\mathbf{p},\mathbf{q}) = \theta(p-k/\alpha)\theta(q-k/\alpha)K(\mathbf{k},\mathbf{p},\mathbf{q}), \quad (C.1b)$$

where $\theta$ is the Heaviside step function and $\alpha$ is a wavenumber cut-off parameter. Then, the Markov version of the regularized the response function equation for $t \geq t'$ becomes

$$\frac{\partial}{\partial t}R_{\mathbf{k}}(t,t') + \sum_{j=0}^{5}\breve{D}_{\mathbf{k}}^{j}(t)R_{\mathbf{k}}(t,t') = \delta(t-t'). \quad (C.2)$$

Here, $\breve{Q}_{\mathbf{k}}^{\prime}$ are defined through equations (5.6) and (5.7) with the replacements $A(\mathbf{k},\mathbf{p},\mathbf{q}) \to \breve{A}(\mathbf{k},\mathbf{p},\mathbf{q})$ and $K(\mathbf{k},\mathbf{p},\mathbf{q}) \to \breve{K}(\mathbf{k},\mathbf{p},\mathbf{q})$. In the original notation the regularized Markov version of response function equation takes the form

$$\frac{\partial}{\partial t} R_{\mathbf{k}}(t,t') + \left[ \nu_0(\mathbf{k})k^2 + \int_{t_0}^{t} ds\{(\tilde{\eta}_{\mathbf{k}}(t,s) + \breve{\pi}_{\mathbf{k}}(t,s))C_{-\mathbf{k}}(t,s)[C_{\mathbf{k}}(t,t)]^{-1}\} \right] R_{\mathbf{k}}(t,t')$$
$$= \delta(t-t') \qquad (C.3)$$

for $t \geq t'$. Here, $\tilde{\eta}_{\mathbf{k}}(t,s)$ is given by equation (4.5a) and $\breve{\pi}_{\mathbf{k}}(t,s)$ by equation (4.10d) with the replacements $A(\mathbf{k},\mathbf{p},\mathbf{q}) \to \breve{A}(\mathbf{k},\mathbf{p},\mathbf{q})$ and $K(\mathbf{k},\mathbf{p},\mathbf{q}) \to \breve{K}(\mathbf{k},\mathbf{p},\mathbf{q})$. In the equations for the triad relaxation times $\breve{Q}_{\mathbf{k}}^{\prime}$ replace $Q_{\mathbf{k}}^{\prime}$ but the single-time covariance equations (4.9) and (5.8) and the mean-field equations (4.7) and (5.12) remain unchanged.

The studies of Frederiksen and Davies (2004) and OF04 suggest that the correct small scale spectral behavior is obtained with a value of $\alpha$ which is essentially universal or only weakly flow dependent.

## 13. Tables

Table 1

| $\Delta t$ | $\hat{v}$ | $C_{\mathbf{k}}(0,0)$ | $\langle \zeta_{\mathbf{k}}(0) \rangle$ | $a$ | $b$ | $U$ (ms$^{-1}$) |
|---|---|---|---|---|---|---|
| 0.21 | $3.378 \times 10^{-5}$ | $\dfrac{0.01 k^2}{a + b k^2}$ | $-10 b h_{\mathbf{k}} C_{\mathbf{k}}(0,0)$ | $4.824 \times 10^4$ | $2.511 \times 10^3$ | 7.5 |

Table 1. Parameters and initial conditions used in calculations and for figures.

## 14. Figures

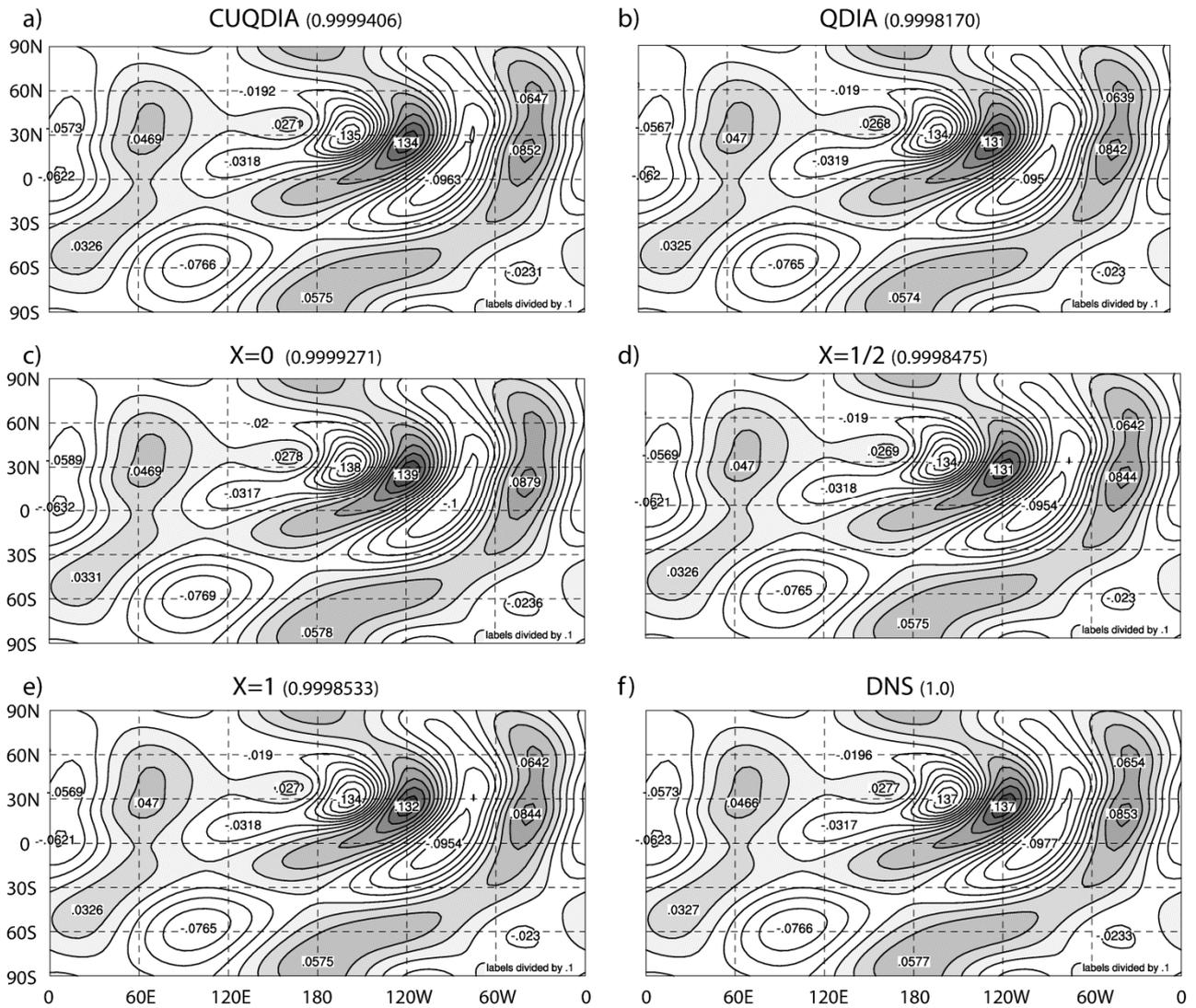

Figure 1. (a) The evolved day 10 CUQDIA, (b) QDIA, (c) MIC with $X = 0$, (d) MIC with $X = \tfrac{1}{2}$, (e) MIC with $X = 1$, and (f) ensemble of DNS Rossby wave streamfunction in non-dimensional units. Multiply values by $\tfrac{1}{4} a_e^2 \Omega^{-1} = 740 \text{ km}^2\text{s}^{-1}$ to convert to units of $\text{km}^2\text{s}^{-1}$.

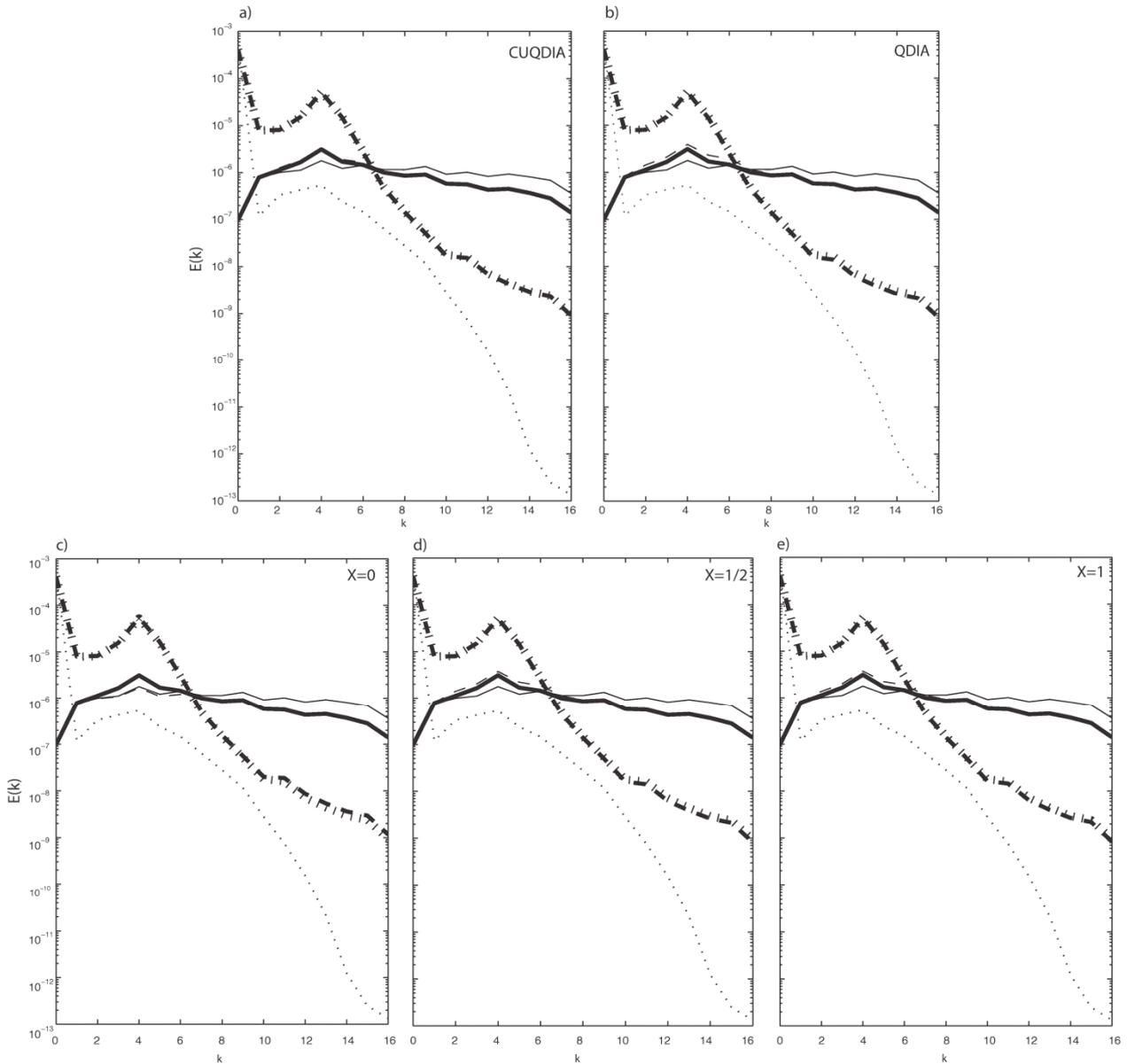

Figure 2. The evolved non-dimensional kinetic energy spectra for the five inhomogeneous closures (a) CUQDIA, (b) QDIA, (c) MIC with $X=0$, (d) MIC with $X=\frac{1}{2}$, (e) MIC with $X=1$, and the ensemble of DNS on each figure part. Show are: initial mean energy (dotted), initial transient energy (thin solid), evolved DNS transient energy (thick solid), evolved DNS mean energy (short wide dashed), evolved closure transient energy (thin dashed) and evolved closure mean energy (thick dashed). Multiply by $\frac{1}{4}a_e^2\Omega^{-2}=5.4\times10^4$ m$^2$s$^{-2}$ to convert to units of m$^2$s$^{-2}$.